\documentclass[sigconf]{acmart}
\usepackage{color,xcolor}
\usepackage{graphicx}
\usepackage{lipsum}

\usepackage{array}
\usepackage{booktabs}
\usepackage{colortbl}
\usepackage{multirow}
\usepackage{float}
\usepackage{caption}
\usepackage{adjustbox}
\usepackage{subcaption}
\usepackage{footnote}
\makesavenoteenv{tabular}
\makesavenoteenv{table}
\usepackage{makecell}
\usepackage{tablefootnote}

\usepackage{amsmath,amsfonts,bm}
\usepackage[super]{nth}

\usepackage{changepage}
\usepackage{extramarks}
\usepackage{xspace}

\usepackage{url}
\usepackage{hyperref}
\hypersetup{colorlinks=True, urlcolor=black}

\usepackage[ruled,vlined]{algorithm2e}
\usepackage{enumitem}
\usepackage{verbatim}
\usepackage{pifont}

\vspace{-3mm}
\copyrightyear{2025}
\acmYear{2025}
\setcopyright{acmlicensed}\acmConference[MM '25]{Proceedings of the 33rd ACM International Conference on Multimedia}{October 27--31, 2025}{Dublin, Ireland}
\acmBooktitle{Proceedings of the 33rd ACM International Conference on Multimedia (MM '25), October 27--31, 2025, Dublin, Ireland}
\acmDOI{10.1145/3746027.3754829}
\acmISBN{979-8-4007-2035-2/2025/10}

\begin{document}

\title{EmoVoice: LLM-based Emotional Text-To-Speech Model with Freestyle Text Prompting}

\author{Guanrou Yang}
  \orcid{0009-0008-3614-1346}
\affiliation{
  \institution{Shanghai Jiao Tong University}
  \city{Shanghai}
  \country{China}
}

\author{Chen Yang}
\orcid{0009-0004-7972-5457}
\affiliation{
  \institution{Shanghai Jiao Tong University}
  \city{Shanghai}
  \country{China}
}

\author{Qian Chen}
\authornote{Corresponding author.}
\orcid{0000-0001-6939-7438}
\affiliation{
  \institution{Tongyi Speech Lab}
  \city{Hangzhou}
  \country{China}
}

\author{Ziyang Ma}
\orcid{0000-0002-8195-3262}
\affiliation{
  \institution{Shanghai Jiao Tong University}
  \city{Shanghai}
  \country{China}
}

\author{Wenxi Chen}
\orcid{0009-0005-8303-930X}
\affiliation{
  \institution{Shanghai Jiao Tong University}
  \city{Shanghai}
  \country{China}
}

\author{Wen Wang}
\orcid{0000-0002-2765-5889}
\affiliation{
  \institution{Tongyi Speech Lab}
  \city{Sunnyvale}
  \country{United States}
}

\author{Tianrui Wang}
\orcid{0000-0002-2765-5889}
\affiliation{
  \institution{Tianjin University}
  \city{Tianjin}
  \country{China}
}

\author{Yifan Yang}
\orcid{0009-0003-0588-1812}
\affiliation{
  \institution{Shanghai Jiao Tong University}
  \city{Shanghai}
  \country{China}
}

\author{Zhikang Niu}
\orcid{0009-0007-1880-7434}
\affiliation{
  \institution{Shanghai Jiao Tong University}
  \city{Shanghai}
  \country{China}
}

\author{Wenrui Liu}
\orcid{0009-0000-5940-5369}
\affiliation{
  \institution{Zhejiang University}
  \city{Hangzhou}
  \country{China}
}

\author{Fan Yu}
\orcid{0009-0005-5958-1131}
\affiliation{
  \institution{Tongyi Speech Lab}
  \city{Hangzhou}
  \country{China}
}

\author{Zhihao Du}
\orcid{0000-0003-3509-9322}
\affiliation{
  \institution{Tongyi Speech Lab}
  \city{Hangzhou}
  \country{China}
}

\author{Zhifu Gao}
\orcid{0009-0008-5691-7324}
\affiliation{
  \institution{Tongyi Speech Lab}
  \city{Hangzhou}
  \country{China}
}

\author{Shiliang Zhang}
\orcid{0000-0003-1718-3686}
\affiliation{
  \institution{Tongyi Speech Lab}
  \city{Hangzhou}
  \country{China}
}

\author{Xie Chen}
\authornotemark[1]
\orcid{0000-0001-7423-617X}
\affiliation{
  \institution{Shanghai Jiao Tong University, Shanghai Innovation Institute}
  \city{Shanghai}
  \country{China}
}

\renewcommand{\shortauthors}{Guanrou Yang et al.}

\begin{abstract}
Human speech goes beyond the mere transfer of information; it is a profound exchange of emotions and a connection between individuals. 
While Text-to-Speech (TTS) models have made huge progress, they still face challenges in controlling the emotional expression in the generated speech. 
In this work, we propose \textbf{EmoVoice}, a novel emotion-controllable TTS model that exploits large language models (LLMs) to enable fine-grained freestyle natural language emotion control, and a phoneme boost variant design that makes the model output phoneme tokens and audio tokens in parallel to enhance content consistency, inspired by chain-of-thought (CoT) and chain-of-modality (CoM) techniques.
Besides, we introduce \textbf{EmoVoice-DB}, a high-quality 40-hour English emotion dataset featuring \textit{expressive speech} and \textit{fine-grained emotion labels} with natural language descriptions.
EmoVoice achieves state-of-the-art performance on the English EmoVoice-DB test set using only synthetic training data, and on the Chinese Secap test set using our in-house data.
We further investigate the reliability of existing emotion evaluation metrics and their alignment with human perceptual preferences, and explore using SOTA multimodal LLMs GPT-4o-audio and Gemini to assess emotional speech. 
Dataset, code, checkpoints and demo samples are available at \url{https://github.com/yanghaha0908/EmoVoice}.



\end{abstract}


\vspace{-4mm} 


\vspace{-8mm} 
\maketitle
\vspace{-4mm} 

\section{Introduction}
Text-to-Speech (TTS) technology has been advancing at an incredible pace. 
However, human communication encompasses not only the conveyance of information but also the nuanced expression of emotions.
We expect to synthesize speech that mirrors the emotional richness and expressiveness of human speech.
Emotion-controllable TTS models play a pivotal role in achieving this goal by generating emotionally authentic and resonant speech tailored to specific requirements.
They hold significant practical value in applications such as virtual assistants and emotional companions. 
However, the domains of emotion-controllable TTS models, emotion datasets, and evaluation metrics are still in their infancy, leaving ample room for further exploration and development.

Prior research efforts on emotional TTS primarily rely on coarse emotion category labels~\cite{ZET-Speech, emomix, emodiff, emospeech}, which are insufficient to comprehensively capture the nuanced emotion states conveyed in speech.
Ideally, we envision a model where fine-grained natural language descriptions can steer the emotional tone of synthesized speech, providing an intuitive and user-friendly interface for emotion control. 
Some studies explore style-controllable TTS using natural language prompts~\cite{prompttts, prompttts2, instructtts,textrolspeech}, but they primarily focus on concrete attributes such as gender, pitch, duration, and energy, rather than addressing abstract emotion characteristics. 

On the other hand, high-quality emotion datasets are extremely scare. Existing real-world open-source emotion speech datasets, as systematically compiled by EmoBox~\cite{emobox}, amount to less than 300 hours in total, despite aggregating data across multiple languages. 
In addition to the limited scale, these datasets are also limited to coarse emotion category labels, speech with indistinct emotions, and inconsistent labeling . They only provide \textit{coarse} emotion category labels, such as a single adjective like ‘happy’ or ‘sad’. 
Some audio samples do not match their assigned emotion labels, sounding indistinguishable from neutral speech or from different emotion categories. Besides, the criteria for emotion classification vary across datasets: some datasets assign labels based on subtle emotional tendencies, while others require clear and explicit expressions, resulting in inconsistencies. Clearly, these datasets are insufficient for training fine-grained emotion-controllable TTS models.

Furthermore, robust emotion evaluation metrics are still lacking. Currently, time-consuming and costly subjective evaluations remain the most accurate way to assess emotion accuracy.
Some works utilize emotion2vec~\cite{emotion2vec} to classify synthesized speech and calculate the accuracy~\cite{emosphere++,lucy,cosyvoice,openomni}, or extract embeddings from both synthesized and ground truth speech and calculate similarity~\cite{emodpo,emopro,emosphere++,laughnow}. However, the emotion resolution of these methods is constrained by that of the evaluation model, and their reliability for fine-grained emotion evaluation remains unproven.

In this paper, we explore the task of fine-grained emotion controllable TTS models with freestyle text prompting, and address limitations of existing works from the perspectives of \textit{model design}, \textit{data}, and \textit{evaluation metrics}. At the model level, we employ an LLM-based TTS model to enhance comprehension of fine-grained emotion descriptions in natural language, harnessing LLM's capabilities in textual semantic understanding and emotion analysis. Our model allows direct input of emotion descriptions into the LLM, using the LLM to comprehend instructions without dedicated prompt encoders required by previous models~\cite{prompttts}. 
Besides, inspired by the CoT~\cite{cot} technique for LLMs and CoM proposed in SpeechGPT~\cite{speechgpt}, our model is designed to output phoneme tokens and audio tokens in parallel, forcing the model to predict pronunciation before generating the final audio tokens, hence boosting the performance.
At the data level, we construct a high-quality synthetic emotion dataset by distilling the current best-performing TTS model, GPT-4o-audio\footnote{\label{gpt-4o-audio}\url{https://platform.openai.com/docs/models/gpt-4o-audio-preview}}, which excels in emotional expressiveness and instruction-following capabilities. 
At the evaluation metric level, we delve into the reliability of existing emotion metrics and their alignment with human preferences, and pioneer the effort to evaluate emotional speech with SOTA multimodal LLMs GPT-4o-audio\footref{gpt-4o-audio} and Gemini\footnote{\url{https://ai.google.dev/gemini-api/docs/models##gemini-2.0-flash}} with comprehensive capabilities. 

Our main contributions can be summarized as follows:
\begin{itemize}[leftmargin=*, topsep=0pt]
\item We propose EmoVoice, a novel fine-grained freestyle text prompt-based emotion-controllable TTS model based on LLMs. 
Inspired by CoT and CoM methods, we propose EmoVoice-PP, a variant of EmoVoice that outputs audio and phoneme tokens in parallel to improve content consistency. 
\item We construct EmoVoice-DB, a high-quality 40-hour English emotion dataset with expressive speech and natural language emotion description labels.
\item EmoVoice achieves SOTA performance on English EmoVoice-DB and Chinese Secap test sets. 
By using synthetic data from CosyVoice for pre-training and from GPT-4o-audio for emotional fine-tuning, we demonstrate the feasibility of training a superior emotional TTS model with only synthetic data.



\item We systematically evaluate the alignment with human preferences for existing speech emotion evaluation metrics and explore using advanced multimodal LLMs to assess emotional speech. 
\end{itemize}

\vspace{-3mm} 
\section{Related work}

\vspace{-1mm}
\subsection{Emotional TTS}
Emotion-controllable TTS models can be broadly categorized into two categories, based on dataset labels: coarse-grained emotion category and fine-grained natural language emotion description.

\noindent\textbf{Coarse-grained emotion category.}
Earlier works mainly use category labels for emotion control, focusing on mainstream categories such as happy, sad, angry, etc. 
Some emotional TTS systems adopt diffusion models~\cite{emodiff,ZET-Speech,emomix,edtts}. Among them, EMODIFF~\cite{emodiff} allows manipulation of emotion intensity through soft-label. EmoMix~\cite{emomix} combines a diffusion probabilistic model with a pre-trained speech emotion recognition model, enabling generation of emotional speech with specified intensity or blended emotions. 
\cite{qitts,fine-grained, emosphere,emospeech} build on the non-autoregressive transformer-based FastSpeech2 model~\cite{fastspeech2, vittts}. For example, EmoSpeech~\cite{emospeech} introduces a conditioning mechanism to address uneven emotion distribution in text, enabling emotion intensity variation across phonemes for improved intonation perception. 
There are also explorations beyond model architecture. For example, Emo-DPO~\cite{emodpo} leverages an LLM-TTS architecture and direct preference optimization to capture subtle emotional nuances. EmoPro~\cite{emopro} proposes a two-stage prompt selection strategy for selecting high-quality prompt speech.


\noindent\textbf{Fine-grained natural language description.}
Using natural language descriptions as emotion labels, such as \textit{Expressing supportive joy and pride in someone’s accomplishment}, is superior to simple categorical tags by capturing complex and nuanced emotional states, allowing for richer emotional expression in speech synthesis.
However, no existing TTS models are specifically designed to control emotions through natural language prompts. Most instead focus on general style control, adjusting attributes including gender, pitch, speed, volume, and emotion.
For example, promptTTS~\cite{prompttts} conditions FastSpeech2 on style representations extracted from prompts with a style encoder. It trains on a custom dataset with five style factors—gender, pitch, speed, volume, and emotion.
InstructTTS~\cite{instructtts} trains a discrete diffusion model~\cite{speechtoken,medic} to generate discrete acoustic features, and builds a Chinese corpus focusing on gender, pitch, speech, and volume.
PromptTTS2\cite{prompttts2} is a diffusion-based variation network modeling voice variability beyond prompts to address the one-to-many problem.
Salle~\cite{textrolspeech} integrates style prompts into the Valle\cite{valle} model.
ControlSpeech~\cite{controlspeech} is capable of simultaneously cloning a speaker’s voice based on an audio prompt and controlling speaking style with a textual style prompt.

PromptStyle~\cite{promptstyle} achieves controllable style transfer through natural language prompts, while ~\cite{controlling} implicitly utilizes emotional semantics in an emotionally rich text as a prompt to control speech.

The most relevant works to ours are CosyVoice~\cite{cosyvoice,cosyvoice2}. CosyVoice-Instruct, applies instruction fine-tuning on CosyVoice to control speaker identity, speaking style, and fine-grained paralinguistic features, based on hundreds of hours of internal data. CosyVoice2 integrates 1500 hours of instructed training data into the base training set, seamlessly integrating the instruction and zero-shot capabilities into a single model.
Our EmoVoice differs from CosyVoice2 in three key aspects. We employ group token modeling, predicting three audio tokens simultaneously at each decoding step, which significantly accelerates training and reduces modeling difficulty. 
Our EmoVoice-PP outputs phoneme tokens in parallel to guide audio token generation, which improves content consistency.
 Our models are trained on emotion-specific data of higher quality.
Besides, OpenAI's GPT-4o-audio remains the most advanced TTS model for natural language instruction control. The latest  GPT-4o-mini-TTS also supports prompting to control various speech characteristics.

\vspace{-5mm}
\subsection{LLM-based TTS}
LLMs have made significant advancements in natural language processing and are expanding their capabilities to speech and vision modalities. While most previous work concentrates on prompting LLMs with audio comprehension abilities, some studies also explore adapting LLMs for the TTS task.
\cite{boosting} presents a comprehensive empirical study on enhancing LLMs with speech generation capabilities by integrating pre-trained Llama/OPT~\cite{llama,opt} with VALL-E, comparing three integration methods.
TTS-Llama~\cite{ttsllama} fine-tunes the Llama3-8B-Instruct with LoRA~\cite{lora}, being the first TTS system to achieve SOTA performance through only PEFT fine-tuning of a text-based LLM.
Spark-TTS~\cite{sparktts} is an innovative system powered by a single-stream speech codec that disentangles semantic and speaker information from speech, utilizing Qwen2.5-0.5B~\cite {qwen2.5} as backbone, integrating TTS capabilities within the text LLM paradigm.
Llasa~\cite{llasa} is initialized from the Llama by expanding vocabulary to incorporate speech tokens, trained using the next-token prediction paradigm. It explores the scaling of compute for TTS task, and proposes a single-layer codec to better align with standard LLMs.
CosyVoice2 upgrades CosyVoice by utilizing the Qwen2.5-0.5B to autoregressively generate speech tokens from text prompts, eliminating the need for a separate text encoder as the LLM is powerful enough to align the text and speech tokens.
Our EmoVoice also adopts Qwen2.5-0.5B as the backbone.
Besides, the Omni series models primarily employ LLM as the backbone to enable end-to-end speech conversation tasks. Mini-Omnis~\cite{mini-omni,mini-omni2}, and SLAM-Omni~\cite{slam-omni} utilize the Qwen2-0.5B as their foundational model, while LLama-Omni~\cite{llama-omni} adopts Llama-3.1-8B-Instruct, and Freeze-Omni~\cite{freeze-omni} employs Qwen2-7B-Instruct3.






\vspace{-4mm} 
\section{Methodology}
\vspace{-1mm}

\subsection{Model}
\begin{figure}[t]
  \centering
  \includegraphics[width=1\linewidth]{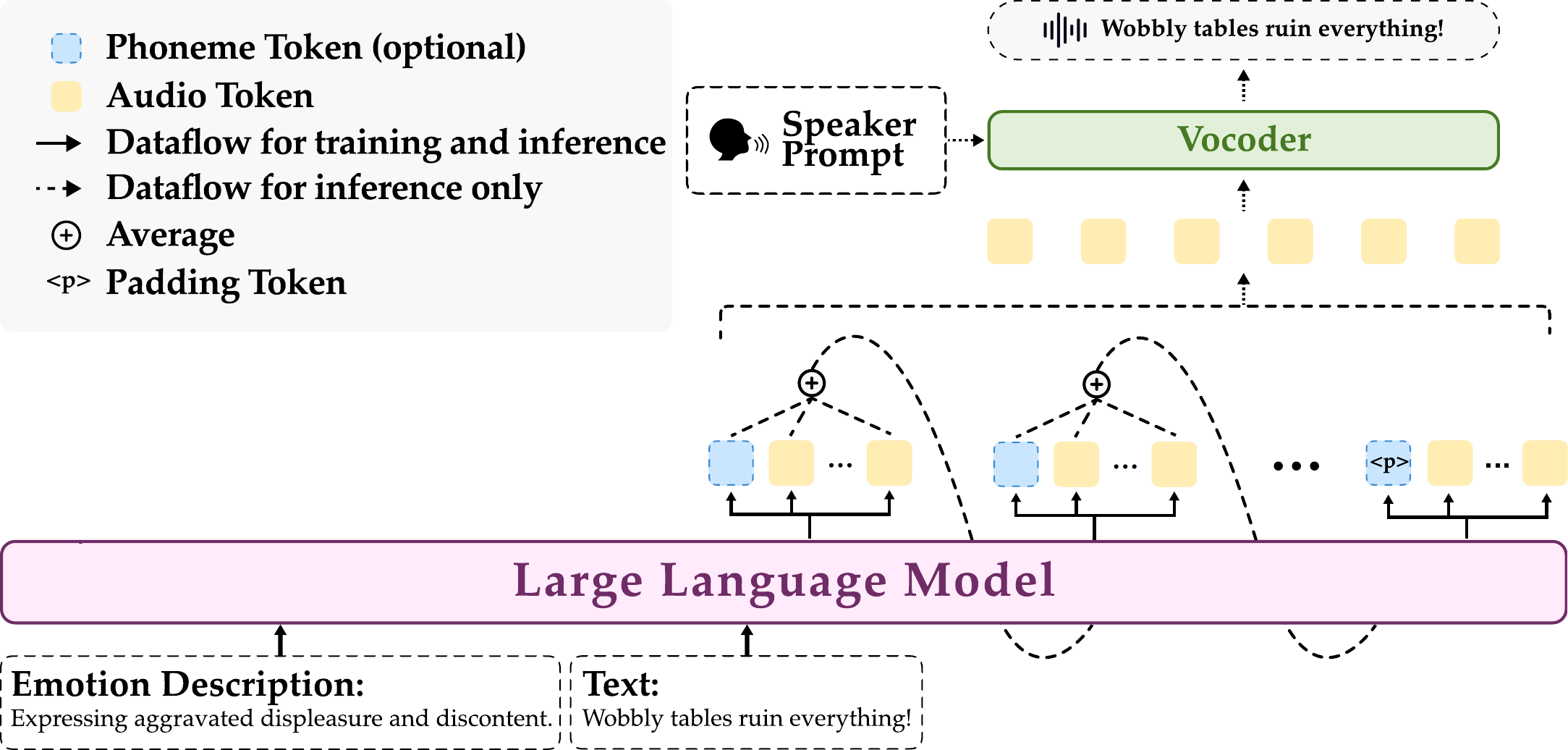}
   \vspace{-4mm}
  \caption{Overview of EmoVoice. LLM receives emotion description and text, autoregressively generating audio token group and phoneme token(optional) in parallel. 
  }
  \label{fig:model}
  \vspace{-5mm} 
\end{figure}

\noindent\textbf{EmoVoice}
The architecture of our proposed model, EmoVoice, is illustrated in Figure \ref{fig:model}. Its backbone is based on Qwen2.5-0.5B, which is a causal pre-trained language model consisting of 24 Transformer layers and approximately 0.49 billion parameters. We initialize the LLM model using the parameters of Qwen2.5-0.5B.
The input consists of pure text, including a fine-grained description of emotion and the text to be generated, organized into the following format:  
  ``\textit{<SYSTEM>: Say this sentence with emotion of <Description>. \textbackslash n <Text>}.''
The input text is tokenized using Qwen2.5-0.5B tokenizer.

EmoVoice autoregressively predicts the 50Hz CosyVoice semantic tokens as speech output, which is ultimately transformed into an audio waveform through the flow matching module and HiFi-GAN~\cite{hifi-gan} vocoder in CosyVoice. 
We extend the original LLM vocabulary \( V_t \) and its embedding space by introducing a new codebook \( V_a \) specifically for audio tokens, forming an expanded vocabulary \( V_j = V_t \cup V_a \). The vocab embedding matrix of the original LLM remains unchanged, while the embeddings for the audio tokens are initialized randomly. At each prediction step, we extract the audio part of the output logits, \( x_a = \text{logits}[..., |V_t|:] \), which represents the predicted distributions for the audio tokens. 

Additionally, we employ semantic group modeling proposed in SLAM-Omni to compress the generated sequence length, which accelerates training and reduces modeling difficulty. At each prediction step, we predict \( G \) semantic tokens, where \( G \) is the group size. Specifically, we use a linear layer to project the audio logits \( L_a \) into group-sized logits \( L_g \), and \( L_g \in \mathbb{R}^{|V_a| \times G}\). Meanwhile, the model’s input at each prediction step is the average embedding value of each semantic token within the group. 
We calculate the cross-entropy loss on the output semantic tokens, following standard practice.



\noindent\textbf{EmoVoice-PP}
EmoVoice-PP is a phoneme boost variant model of EmoVoice. It adopts parallel audio-phoneme modeling on the output side, simultaneously predicting both semantic and phoneme tokens. 
Ground truth phoneme sequences are extracted by the Phonemizer\cite{phonemizer} toolkit for teacher forcing training.
In the inference stage, due to the lower token rate of phoneme tokens(\textasciitilde 11Hz) compared to audio tokens( \textasciitilde 17Hz), they are predicted earlier and serve as intermediate supervision signals to guide the final generation of audio tokens, which improves content consistency.
Since the Qwen2.5-0.5B tokenizer's vocabulary does not include phonemes, we first add each phoneme as a new token to its vocabulary, resulting in a modified vocabulary \( V_t' \), with the embeddings corresponding to the phoneme tokens initialized randomly. At each prediction step, we separately extract the audio and phoneme parts from the output logits: \( x_a = \text{logits}[..., |V_t'|:] \) and \( x_p = \text{logits}[..., :|V_t'|] \), which represent the predicted distributions for the audio and phoneme tokens, respectively. Meanwhile, the model’s input at each prediction step is the average embedding value of all semantic tokens within the group alongwith the phoneme token.


\vspace{-3mm} 
\subsection{Training Pipeline}
The training of EmoVoice comprises two sequential phases. In the first phase, we pre-train the model on standard TTS training data to develop a robust and stable TTS model. The input text is formatted as: "\textit{<SYSTEM>: Say this sentence. \textbackslash n <Text>}." The second phase involves fine-tuning with a limited set of instruction data composed of text, natural language emotion descriptions, and emotionally expressive speech, establishing an emotion-controllable TTS model. The input text follows the structured format: \textit{<SYSTEM>: Say this sentence with emotion of <Description>. \textbackslash n <Text>}."



\vspace{-3mm} 
\section{EmoVoice-DB}
\vspace{-1mm} 

\begin{table*}[ht]
\vspace{-3mm}
    \centering
    \caption{Statistics and Examples of EmoVoice-DB Dataset}
    \vspace{-3mm}
    \label{tab:emovoice-db}
    \resizebox{\linewidth}{!}{%
    \begin{tabular}{lccll}
        \toprule
        \textbf{Emotion} & \textbf{Count} & \textbf{Duration (h)} & \textbf{Text Example} & \textbf{Emotion Description Example} \\
        \midrule
        Angry & 3486 & 5.76 & Wobbly tables ruin everything! & Expressing aggravated displeasure and discontent. \\
        Happy & 3269 & 6.02 & You did an AMAZING job on the presentation! & Expressing supportive joy and pride in someone's accomplishment. \\
        Sad & 3174 & 6.94 & Cracked earth stretches for miles, nothing GREEN to soothe the eye. & Conveying a pervasive sense of desolation and despair. \\
        Surprised & 3072 & 5.67 & The curtain rose without warning, revealing a stage of impossible colors and shapes. & Evoking an excited and bewildered wonder in a rising, quickened cadence. \\
        Fearful & 2961 & 5.52 & Moonlight glinted off the knife, casting shadows that DANCED like ghosts on the walls. & Emanating a chilling foreboding, underscored by a quivering voice. \\
        Disgusted & 2950 & 5.59 & How could anyone EVER think that brown and pink match! & Expressing a moment of incredulous disdain and distaste. \\
        Neutral & 3188 & 4.95 & Leaves rustled in the evening breeze, swaying gently to unseen rhythms. & Emanating a peaceful, contemplative atmosphere. \\
        \midrule
        \textbf{Sum} & \textbf{22100} & \textbf{40.45} & & \\
        \bottomrule
    \end{tabular}
    }
    \vspace{-3mm}
\end{table*}

\subsection{Overview of EmoVoice-DB}
We introduce EmoVoice-DB, an unprecedented, high-quality English emotional speech dataset featuring fine-grained emotion labels expressed through natural language descriptions. This dataset contains over 20,000 emotionally expressive speech samples, each annotated with detailed and precise emotional descriptions, totaling approximately 40 hours of audio. EmoVoice-DB is built using synthetic data generated by the powerful GPT-4o models.

The EmoVoice-DB dataset spans seven core emotion categories— angry, happy, sad, surprised, disgusted, fearful, and neutral—with a balanced distribution of samples across all emotional classes. It features a diverse range of textual content, including novel excerpts, dialogue, and observational phrases. Additionally, the dataset includes speech samples of five distinct speaker timbres, enhancing the diversity of vocal expression. All emotional speech samples are synthesized using the advanced GPT-4o-audio model, ensuring precise emotional control, strong expressiveness, and human-level naturalness. A detailed statistical overview and examples of the dataset are provided in Table \ref{tab:emovoice-db}. EmoVoice-DB provides a valuable resource for advancing research in fields such as emotional speech synthesis, speech emotion recognition, and emotion analysis. 

\vspace{-2mm} 
\subsection{Dataset Construction Process}

EmoVoice-DB construction follows a systematic three-step pipeline. 

\noindent\textbf{Step 1: Generating text and emotional descriptions.} In the first step, pairs of texts and corresponding emotional descriptions are generated using the GPT-4o model. For each emotional category, the model is prompted to produce text that aligns with the specified emotion and a detailed emotional description that captures the intended sentiment. The text generation process is guided by a strict set of constraints to ensure quality and consistency. Each utterance is required to be between 15 and 25 words in length, corresponding to 5–10 seconds of speech. The text should contain rich sensory and emotional details while avoiding clichéd or overused patterns. To enrich the diversity and depth of the textual content, the texts are evenly distributed across three distinct categories: vivid descriptive sentences in the form of novel prose, e.g., \textit{``Wind whispered through the parched cornstalks, its voice fraying like worn silk.”}, emotionally charged dialog excerpts representing natural spoken lines, e.g., \textit{``I’ve asked you three times! Why is the door still locked?”}, observational phrases offering subtle situational commentary, e.g., \textit{``Rain taps the window like it’s bruising the glass—rhythmic, insistent, all night.”} To guide prosodic emphasis in the synthesized speech, up to two words per sentence could be capitalized to indicate stress. We also impose specific constraints on the emotional description. Each description is formatted as a present participle verb phrase, e.g., \textit{``Conveying a contagious, joyful atmosphere.”}, and should focus on vocal affect, excluding contextual references or specific events. In addition, descriptions are restricted to a single sentence, avoiding overly brief or single-word responses that lack sufficient details.

\noindent\textbf{Step 2: Generating emotion speech.}
In the second step, emotional speech samples are generated by prompting GPT-4o-audio model using both text and emotion descriptions constructed earlier. The model is instructed to synthesize speech using prompts such as: \textit{Repeat this sentence with the emotion of <Description>: <Text>}. It evenly utilizes five distinct speaker voices to ensure diversity in vocal timbre and speaking style.

\noindent\textbf{Step 3: Post-processing.}
The final step involves post-processing and filtering. While GPT-4o-audio is capable of high-fidelity synthesis, occasional issues such as misread words, omissions, or improvisational deviations from the script are observed in the synthesized speech. To ensure transcriptional fidelity, we compute WER between the generated speech and the intended transcript. Samples with high WER are filtered out, resulting in a final dataset that preserves both emotional expressiveness and linguistic precision.

\vspace{-4mm} 
\section{Experimental Setup}

\noindent \textbf{Datasets.}
Since our training process consists of two stages of pretraining a standard TTS model and fine-tuning it for emotional speech synthesis, we utilize both neutral TTS training data and emotional speech data. Notably, our English model relies entirely on synthetic training data. 

For pretraining stage, we employ the VoiceAssistant~\cite{mini-omni} dataset for the English model and the Belle~\cite{belle} dataset for the Chinese model. VoiceAssistant and Belle are initially text dialogue datasets in English and Chinese, respectively. SLAM-Omni utilizes them to construct synthesized speech dialogue datasets.
It employs the text-to-token LLM from CosyVoice to generate semantic tokens, with enriched vocal timbre.
For training, we use only the response portions of the dialogues, containing 3,234 and 6,418 hours of speech from 
VoiceAssistant and Belle respectively.

For fine-tuning stage, we employ two datasets for the English model: a synthetic emotional speech dataset by GPT-4o-audio with categorical labels, LAION's Got Talent, and our proposed EmoVoice-DB, which features freestyle natural language descriptions of emotions.
The LAION's Got Talent dataset consists of voice acting samples that span a broad range of emotions, vocal bursts, topics, and content.
The emotional labels combine multiple synonyms to describe a specific emotion, such as "intense anger, rage, fury, hatred, and annoyance." These descriptions are fixed and finite, resembling emotion classification tags more than natural language descriptions. We filter the dataset to include only English sentences with no accent, resulting in 44.4k samples, amounting to about 200 hours of speech. This filtered data is entirely used as the training set. 
Additionally, we randomly sample 100 utterances per emotion category from the EmoVoice-DB dataset for the test set and 50 utterances for the validation set, with the remaining audio serving as the training set. 
We utilize an internal industrial training dataset for the Chinese model and adopt the Secap~\cite{secap} test set for evaluation, which contains a total of 600 audio samples. For all emotional fine-tuning data, we extract 50 Hz CosyVoice semantic tokens from each utterance. The resulting text-token pairs are directly used as input training data.

\noindent \textbf{Training and Inference Details.}
In the training process, EmoVoice (0.5B) is initialized with the pre-trained Qwen2.5-0.5B model and train all parameters. The complete architecture, comprising the LLM and grouped decoding linear layer, contains 550M trainable parameters. EmoVoice(1.5B) utilizes Qwen2.5-1.5B of 28 Transformer layers with 1.54B parameters.
We use the AdamW~\cite{adamw} optimizer with $\beta=(0.9,0.999)$ and zero weight decay. The semantic token group size is set to 3, which enables 2.64x faster training than standard single-layer audio token output, with lower WER and higher content accuracy.
During the standard TTS pre-training phase, the learning rate increases linearly from 0 to a peak rate of 1e-4 over the first 1,000 steps, and then decays linearly to zero during the remaining training time. In the emotional TTS fine-tuning stage, the learning rate scheduler follows the same pattern, but with the peak rate set to 1e-5. We conduct our experiments on 4 80GB A800 GPUs and set the batch size to 6.
In the inference process, we use greedy search decoding algorithm with a repetition penalty of 1.2. We employ flow matching model and HiFi-GAN vocoder of CosyVoice to convert semantic tokens into audio waveforms. We pick a different neutral-emotion speech sample of the same speaker with the ground-truth speech, as the prompt speech for the flow matching module to control the timbre of the synthesized audio.


\noindent \textbf{Evaluation Metrics.} 
We evaluate the content consistency of synthesized speech using the Word Error Rate (WER), with transcription results obtained from the Whisper large-v3~\cite{whisper} model.
To assess emotional expressiveness, we adopt two metrics: Emotion Similarity and Recall Rate. Emotion Similarity is computed by extracting emotion embeddings from the synthesized speech using emotion2vec, and calculating the cosine similarity between these embeddings and those from the ground-truth speech.
For Recall Rate, we also leverage emotion2vec to perform emotion classification on the generated speech. Due to insufficient training data and consequently low recognition accuracy of emotion2vec for the categories disgusted, fearful, and surprised, these emotions are excluded from the evaluation. 
Recall is computed as the proportion of correctly classified audio samples within each emotion category, and the final Recall Rate is the average value across all selected categories.
In addition, we evaluate the naturalness and perceptual audio quality of the synthetic speech using the UTMOS score~\cite{utmos}.

\vspace{-2mm} 
\section{Results and Abalation}

\subsection{Main Results}
\subsubsection{Objective evaluation results}

\begin{table}[h]
\vspace{-3mm}
\centering
\caption{Performance comparison of EmoVoice and other emotion-controllable TTS models in terms of WER, emotion similarity, emotion classification recall rate, and UTMOS on EmoVoice-DB test set.}
\vspace{-2mm}
\label{tab:comparison}
\resizebox{\columnwidth}{!}{
\begin{tabular}{lcccc}
\toprule[1pt]
Model          & WER$\downarrow$  & Emo\_Sim$\uparrow$ & Recall$\uparrow$ & UTMOS$\uparrow$ \\ 
\midrule
PromptStyle~\cite{promptstyle}    & 15.51 & 0.8717  & 0.313  & 3.59 \\
PromptTTS~\cite{prompttts}      & \textbf{2.11}  & 0.8709  & 0.291  & 4.32 \\
CosyVoice~\cite{cosyvoice}     & 3.61  & 0.8889  & 0.329  & 4.33 \\
CosyVoice2~\cite{cosyvoice2}     & 3.61  & 0.8647  & 0.37  & \textbf{4.42} \\
EmoVoice(0.5B)           & 2.73  & 0.9100  & 0.395  & 4.36 \\
\textbf{EmoVoice(1.5B)}     & \underline{2.62}   & \textbf{0.9118}  & \textbf{0.424}  & 4.35 \\ \midrule
GPT-4o-mini-tts  & 2.40   & 0.9168  & 0.456  & 4.06 \\
GPT-4o-audio(GT)  & 2.01  & /       & 0.525  & 3.74 \\ 
\bottomrule[1pt]
\end{tabular}
}
\vspace{-3mm}
\end{table}



The experimental results in Table \ref{tab:comparison} demonstrate that our proposed model, EmoVoice, achieves competitive performance across all evaluated metrics. Compared with existing emotional speech synthesis models, EmoVoice(1.5B) achieves the most balanced overall performance, with highest emotion similarity of 0.9118, and highest recall rate of 0.424, while maintaining low WER of 2.62 and high UTMOS score of 4.345. It rivals the powerful GPT-4o-mini-tts in content consistency and emotional expressiveness, while surpassing it in terms of speech quality. 
These results indicate that EmoVoice not only preserves linguistic accuracy and delivers high perceptual quality, but also excels in producing emotionally rich speech that faithfully represents the intended emotional directives.
GPT-4o-audio serves as an upper bound reference rather than a comparable model. 

\subsubsection{Subjective evaluation results}

\begin{table}[ht]
\vspace{-3mm}
\centering
\caption{Comparison of Mean Opinion Score (MOS) for emotional expressiveness and instruction-following capabilities of different emotion-controllable TTS models.}
 \vspace{-2mm}
\label{tab:mos}
\begin{tabular}{lccc}
\toprule
Model & \textbf{MOS}$\uparrow$ & Emo\_Sim$\uparrow$ & Recall$\uparrow$  \\
\midrule
PromptTTS~\cite{prompttts}  & 1.415 & 0.8709 & 0.291 \\
CosyVoice2~\cite{cosyvoice2} & 2.138 & 0.8647 & 0.371 \\
EmoVoice(0.5B)     & 3.163 & 0.9100 & 0.395 \\
\textbf{EmoVoice(1.5B)} & \textbf{3.507} & \textbf{0.9118} & \textbf{0.424} \\
\midrule
GPT-4o-mini-tts  & 3.598 & 0.9168 & 0.456 \\
GPT-4o-audio(GT) & 4.350 & / & 0.525 \\
\bottomrule
\end{tabular}
 \vspace{-3mm}
\end{table}
Given the inherently subjective and abstract nature of emotional expressiveness, objective metrics can only provide limited insight into model performance. Therefore, we conduct a subjective assessment to better align evaluations with human perception. Specifically, we randomly sample 10 utterances form each of six emotion categories from the test set, totaling 60 audio samples. These samples are evaluated by 30 distinct participants, each of whom rates 10 random groups of speech produced by six representative models. The evaluation focuses on two key aspects: the overall emotional expressiveness of the synthesized speech and its adherence to the given emotional instructions. Ratings are scored on a scale of 1 to 5, with increments of 0.5.

The results presented in Table \ref{tab:mos} reveal that the MOS trends align closely with the objective emotion metrics, validating the consistency of our findings. Notably, the GPT-4o series models exhibit the strongest performance, with GPT-4o-audio achieving the highest MOS of 4.35, reflecting its ability to produce highly compelling and emotionally rich speech. Our EmoVoice models demonstrate competitive performance, surpassing both CosyVoice2 and PromptTTS by a significant margin. Specifically, the EmoVoice(1.5B) achieves a MOS of 3.507, closely approaching the performance of GPT-4o-mini-tts of 3.598 and outperforming CosyVoice2 of 2.138 and PromptTTS of 1.415, reaffirming the effectiveness of our approach in producing emotionally expressive and instruction-aligned speech.

\subsubsection{Chinese model results}
\begin{table}[h]
 \vspace{-3mm}
\centering
\caption{Performance comparison of EmoVoice and other emotion-controllable TTS models on Secap test set.}
\label{tab:chinese}
 \vspace{-2mm}
\resizebox{\columnwidth}{!}{
\begin{tabular}{lcccc}
\toprule
Model             & WER$\downarrow$   & Emo\_Sim$\uparrow$ & Recall$\uparrow$  &  UTMOS$\uparrow$  \\
\midrule
GPT-4o-tts-mini    & 10.34     &   0.6184       &  0.305      &   \textbf{3.42}     \\
GPT-4o-audio   &  14.83    &   0.6497       &    0.351  &   3.25     \\

CosyVoice2~\cite{cosyvoice2}        & 9.13 & 0.7811   & 0.403  & 2.75 \\
EmoVoice          & 7.64 & 0.7933   & 0.411  & 3.19 \\
\textbf{EmoVoice-PP}     & \textbf{7.60}  & \textbf{0.7939}   & \textbf{0.434}  & 3.20 \\
\midrule
Secap(GT)           & 6.62 & /        & 0.800    & 2.78  \\
\bottomrule
\end{tabular}
}
 \vspace{-3mm}
\end{table}

Table \ref{tab:chinese} presents a comprehensive comparison of different emotional TTS models evaluated on the Secap test set. Due to the lack of Chinese language support in PromptStyle, PromptTTS, and CosyVoice-Instruct, these models are excluded from the evaluation. The ground truth speech samples of the Secap test set are included for reference, providing an upper bound for performance assessment.

Among all the evaluated models, EmoVoice-PP performs the best across all major evaluation metrics.
It achieves the lowest WER of 7.6, indicating high intelligibility that closely approaches the upper bound of 6.62. Besides, it obtains the highest emotion similarity score of 0.7939 and the recall rate of 0.434, indicating strong capability in conveying emotional nuance. 
However, due to the limited amount of emotion-specific fine-tuning data, a noticeable gap still exists between EmoVoice-PP and the human reference in emotional expressiveness, suggesting ample room for further refinement. The baseline EmoVoice model follows closely behind EmoVoice-PP.

GPT-series models exhibit relatively poor performance on the Chinese test set. Pronounced issues such as unnatural prosody and accented timbre are observed. During inference, we manually select speakers that produce more native-like Chinese pronunciation and fewer accent-related artifacts, ensuring reliable evaluation of emotional performance. Despite this effort, these models still lag across all emotion metrics, underscoring the current limitations of GPT-4o models in generating emotionally natural and expressive speech in Chinese.

\vspace{-3mm} 
\subsection{Abalation Study}
\begin{figure}[t]
  \centering
  \includegraphics[width=\linewidth]{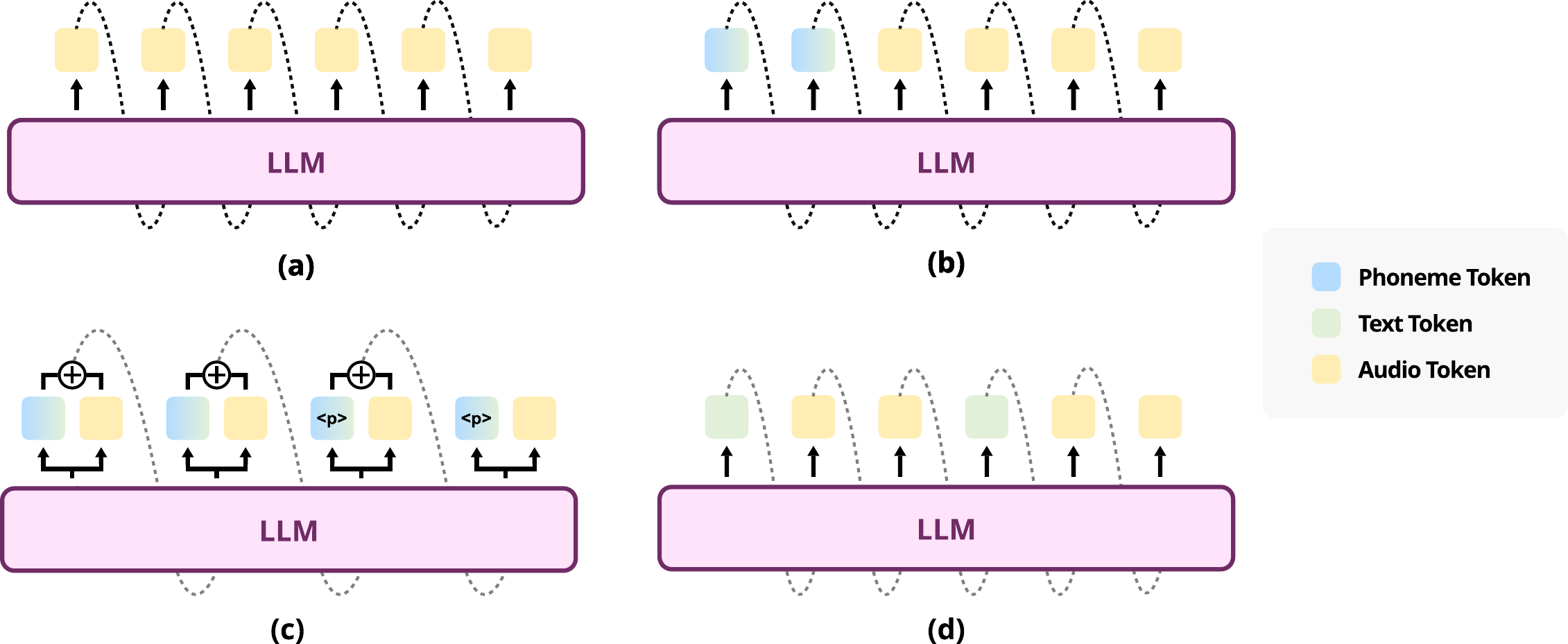}
   \vspace{-3mm}
  \caption{EmoVoice variants with different output structures. (a) Output audio tokens only. (b) Sequential output of phoneme/text tokens followed by audio tokens in a single stream. (c) Parallel output of phoneme/text tokens and audio tokens. (d) Interleaved output of text and audio tokens.}
  \label{fig:output_structure}
  \vspace{-7mm} 
\end{figure}

\subsubsection{Different output-side structures}
\begin{table*}[ht]
 \vspace{-3mm}
\centering
\caption{Performance comparison of EmoVoice variants with different output structures on EmoVoice-DB testset. Both Fine-tuned Emotional TTS and Pre-trained TTS are evaluated. WER of the Pre-trained TTS is also computed on \textit{test-en} of Seed-TTS~\cite{seedtts}.
}

 \vspace{-2mm}
\label{tab:output_structure}
\resizebox{\linewidth}{!}{
\begin{tabular}{llcccccccc}
\toprule[1pt]
\multirow{2}{*}{Model} & \multirow{2}{*}{Output Structure} &\multicolumn{4}{c}{Fine-tuned Emotional TTS} & \multicolumn{4}{c}{Pre-trained TTS} \\
\cmidrule(lr){3-6 } \cmidrule(lr){7-10}
& & WER$\downarrow$&Emo\_Sim$\uparrow$ & Recall$\uparrow$ &  UTMOS$\uparrow$ & WER$\downarrow$  &Emo\_Sim$\uparrow$ & Recall$\uparrow$  &  UTMOS$\uparrow$ \\
\midrule
EmoVoice-I & Interleaved Text and Audio Tokens  &6.91 & 0.9108 & 0.376 & 4.27 & 7.07/4.37 & 0.8727 & 0.303 & 4.39 \\
EmoVoice-SP & Serial Phoneme to Audio Tokens & 3.64 & 0.9079 & 0.383 & 4.31 & 5.57/4.06 & 0.8713 & 0.298 & 4.40 \\
EmoVoice-ST & Serial Text to Audio Tokens &  3.53 & 0.9087 & 0.398 & 4.35 & 5.40/3.43  & 0.8747 & 0.303 & 4.40 \\
EmoVoice-PT & Parallel Text and Audio Tokens & 3.42 & 0.9088 & \textbf{0.405} & \textbf{4.37} & 5.63/3.47 & 0.8711 & 0.291 & 4.40 \\
EmoVoice-PP & Parallel Phoneme and Audio Tokens &3.06 & \textbf{0.9115} & 0.379 & 4.35 & \textbf{3.94/3.11} & 0.8704 & 0.298 & 4.40 \\
EmoVoice & Audio Tokens only& \textbf{2.73} & 0.9100 & 0.395 & 4.36 & 4.73/3.31 & 0.8710 & 0.299 & 4.40 \\
\bottomrule[1pt]
\end{tabular}
}
 \vspace{-2mm}
\end{table*}


We explore optimizing the output-side structure of the EmoVoice model, comparing six different architectures, including standard single-output stream and parallel output streams, as shown in Figure \ref{fig:output_structure}. 
We investigate utilizing text and phonemes as auxiliary information to guide the generation of audio tokens. 
The baseline EmoVoice model outputs a pure audio token sequence.
The EmoVoice-I model produces an interleaved sequence of text and audio tokens, mixed in a predefined ratio of 12:36, a structure commonly employed in streaming TTS models~\cite{cosyvoice2,interleaved}. 
Inspired by CoT and MoT techniques, we experiment with first predicting text or phoneme tokens as an auxiliary intermediate thought process before generating the target audio token within a single output stream. These models are denoted as EmoVoice-ST(Serial Text) and EmoVoice-SP(Serial Phoneme), respectively. 
Besides, we explore using an additional prediction head in the output structure to produce text or phoneme tokens simultaneously. 
EmoVoice-PT(Parallel Text) and EmoVoice-PP(Parallel Phoneme) represent models that parallelly predict audio tokens alongside text tokens and phoneme tokens, respectively.

As shown in Table \ref{tab:output_structure}, emotion similarities and recall rates of these models are nearly comparable, with EmoVoice-PP and EmoVoice-PT slightly outperforming the others, achieving the highest emotion similarity of 0.9115 and recall rate of 0.405. 
The minimal differences in emotional metrics across the models may be attributed to universally high performance. The potential insensitivity of emotion similarity and recall rate metrics may also obscure subtle differences in emotional performance.
Besides, all models exhibit strong perceptual quality in their synthesized speech with consistently high UTMOS scores.
However, notable variations in WER results are observed across the models.
EmoVoice and EmoVoice-PP demonstrate the best performance, with WERs of 2.73 and 3.06, respectively.
The interleaved structure shows relatively high WER, while the remaining models fall within a moderate range of results. 
In the TTS pretraining stage, the EmoVoice-PP model achieves the lowest WER of 3.94 and 3.11 on the EmoVoice-DB and Seed-TTS test-en datasets, respectively, markedly outperforming other models. This highlights the effectiveness of the parallel phoneme sequence output in boosting content consistency of the TTS model.

\subsubsection{Results on hard-case test set}

\begin{table}[h]
\vspace{-2mm}
\centering
\caption{Performance comparison of EmoVoice variants with different output structures on English hard-case testset.}
 \vspace{-2mm}
\label{tab:hard-case}
\resizebox{\columnwidth}{!}{
\begin{tabular}{lcccc}
\toprule[1pt]
Model &  WER$\downarrow$ &Emo\_Sim$\uparrow$ & Recall$\uparrow$  &  UTMOS$\uparrow$ \\
\midrule
EmoVoice        & 18.07 & 0.9084  & 0.23  &  4.23      \\
EmoVoice-PT    & 16.34 & \textbf{0.9154}  & 0.26  &   \textbf{4.32}     \\
\textbf{EmoVoice-PP} & \textbf{11.68} & 0.9100    & \textbf{0.27}   &  4.29      \\
\midrule
GPT-4o-audio(GT)    & 3.95 & / & 0.502  &  3.68      \\
\bottomrule[1pt]
\end{tabular}
}
\vspace{-3mm}
\end{table}

As shown in Table \ref{tab:output_structure}, the performance of different EmoVoice variants models on the EmoVoice-DB test set exhibits minor differences, making it challenging to distinguish their relative strengths and weaknesses. 
To address this, we conduct an extra evaluation experiment on a more challenging hard-case test set, intentionally designed to include difficult sentences, such as tongue twisters, rare words, technical terms, etc.
In this experiment, we focus on the top three models from Table \ref{tab:output_structure}: EmoVoice, EmoVoice-PP, and EmoVoice-PT.

Table \ref{tab:hard-case} shows general poor results, due to inherent pronunciation difficulty of the test set, coupled with the inconsistency between the semantic content of text and the specified emotional instructions.
However, the EmoVoice-PP outperforms the other two models by a noticeable margin in WER, emotion similarity, and recall rate, underscoring the positive impact of incorporating parallel phoneme guidance on model performance. 
Besides, text guidance also brings certain improvement, though less significant compared to phoneme.

\subsubsection{Scaling LLM size}


\begin{table}[ht]
\vspace{-3mm}
\centering
\caption{Effect of scaling LLM (comparing 0.5B and 1.5B) on model performance. WER is computed for both Fine-tuned emotional TTS/Pre-trained TTS models.}
\vspace{-2mm}
\label{tab:scaling}
\resizebox{\columnwidth}{!}{
\begin{tabular}{lccccc}
\toprule[1pt]
Model & Size & WER$\downarrow$ & Emo\_Sim$\uparrow$ & Recall$\uparrow$ & UTMOS$\uparrow$ \\
\midrule
\multirow{2}{*}{EmoVoice} & 0.5B & 2.73/4.73 & 0.9100 & 0.395 & \textbf{4.36} \\
& \textbf{1.5B} & \textbf{2.62/3.83} & \textbf{0.9118} & \textbf{0.424} & 4.35 \\
\bottomrule[1pt]
\end{tabular}
}
\vspace{-3mm}
\end{table}

Given the well-established scaling law in LLMs, we explore how scaling the parameter size of the LLM backbone affects the model's performance on the emotional TTS task. Specifically, we compare two different sizes of 0.5B and 1.5B parameters, using the qwen2.5-0.5B and qwen2.5-1.5B models respectively.

As shown in Tabel \ref{tab:scaling}, in the pre-training phase, we observe an obvious reduction of WER when scaling up the model. The 1.5B model achieves a WER of 3.83\% with a relative reduction of 19.03\% compared to the 4.73\% WER of the 0.5B model, indicating enhanced content consistency.
During the emotion fine-tuning phase, the recall rate rises from 0.395 to 0.424 and the WER reduces from 2.73\% to 2.62\%. Emotion similarity and UTMOS remain largely consistent. Additionally, subjective MOS score in Table \ref{tab:mos} improves from 3.163 to 3.507. 
Overall, scaling up the LLM effectively enhances emotional expressiveness and content consistency.
However, it incurs substantial additional computational costs and training time, with improvements constrained by the limited emotion dataset size.


\subsubsection{LLM Initialization}


\begin{table}[ht]
\vspace{-3mm}
\centering
\caption{Impact of using LLM parameters for initialization on model performance. WER is computed for both Fine-tuned emotional TTS/Pre-trained TTS models.}
\vspace{-2mm}
\label{tab:random_init}
\resizebox{\columnwidth}{!}{
\begin{tabular}{lccccc}
\toprule[1pt]
Model & LLM\_Init & WER$\downarrow$ & Emo\_Sim$\uparrow$ & Recall$\uparrow$ & UTMOS$\uparrow$ \\
\midrule
\multirow{2}{*}{EmoVoice} & \ding{55} & 6.16/7.35 & 0.9033 & 0.387 & 4.36 \\
& \ding{51} & \textbf{2.73/4.73} & \textbf{0.9100} & \textbf{0.395} & 4.36 \\
\midrule
\multirow{2}{*}{EmoVoice-PP} & \ding{55} & 6.06/7.10 & 0.8986 & 0.352 & \textbf{4.37} \\
& \ding{51} & \textbf{3.06/3.94} & \textbf{0.9115} & \textbf{0.379} & 4.35 \\
\midrule
\multirow{2}{*}{EmoVoice-PT} & \ding{55} & 7.87/9.71 & 0.8981 & 0.353 & 4.35 \\
& \ding{51} & \textbf{3.42/5.63} & \textbf{0.9088} & \textbf{0.405} & \textbf{4.37} \\
\bottomrule[1pt]
\end{tabular}
}
\vspace{-3mm}
\end{table}
While several TTS models adopt LLMs as their backbone and utilize their parameters for initialization, the specific impact of LLMs on emotion-controllable TTS tasks remains underexplored. To bridge this gap, we conduct a comparative analysis to evaluate the influence of LLM with different EmoVoice variants. Specifically, we keep the backbone architecture consistent, with the only variation being whether LLM parameters are used for initialization.

The experimental results presented in Table \ref{tab:random_init} reveal that LLM initialization brings significant benefits, consistently improving performance across all three model variants. 
Models initialized with LLM parameters achieve a much lower WER in both fine-tuned emotional models and pre-trained TTS models, underscoring the effectiveness of LLMs in aligning textual and acoustic representations. 
Besides, LLM initialization also enhances the ability to convey emotions, as evidenced by consistent gains in emotion similarity and recall rate. 
These findings highlight that the advanced capabilities of LLMs in language understanding and sentiment comprehension are effectively transferred to the TTS framework, elevating both content accuracy and emotional expressiveness.

\subsubsection{Data Augmentation}



\begin{table}[ht]
 \vspace{-3mm}
\centering
\caption{Impact of emotion description augmentation on EmoVoice-DB training data for EmoVoice model.}
 \vspace{-2mm}
\label{tab:augmentation}
\resizebox{\columnwidth}{!}{
\begin{tabular}{lccccc}
\toprule
Dataset & Model & WER$\downarrow$  & Emo\_Sim$\uparrow$  & Recall$\uparrow$ & UTMOS$\uparrow$ \\
\midrule
EmoVoice-DB & \multirow{2}{*}{EmoVoice} & 3.83 & 0.9089 & \textbf{0.402} & 4.35 \\
\qquad + Description Augmentation &  & \textbf{2.73} & \textbf{0.9100} & 0.395 & \textbf{4.36} \\
\bottomrule
\end{tabular}
}
 \vspace{-3mm}
\end{table}

Given the inherent variability in natural language descriptions compared to fixed emotion category labels, freestyle emotional descriptions theoretically possess infinite variations. A single emotional state can be articulated through numerous linguistic expressions. 
To address this variability and enhance the currently limited EmoVoice-DB dataset, we introduces a data augmentation strategy based on emotional description rewriting. 
Specifically, we leverage GPT-4o to rephrase emotional descriptions while maintaining the original meanings. For each data entry, we generate two rephrased versions, resulting in three semantically equivalent but lexically diverse descriptions per emotional speech sample.
We hope to improve the model’s semantic understanding of emotion descriptions and enhance its robustness.

We compare the performance of EmoVoice trained on the original EmoVoice-DB with that trained on the augmented version with enhanced descriptions. 
As shown in Table \ref{tab:augmentation}, emotional description label augmentation leads to clear performance gains. The WER is reduced from 3.83 to 2.73, with a notable relative improvement of 28.72\%, suggesting the augmented model produces more accurate and coherent outputs. 
The model’s exposure to diverse emotional prompts enhances understanding of emotional semantics and mitigates overfitting to a single expression pattern, thereby improving robustness, generalization ability, and accuracy. 
Besides, it facilitates better alignment between textual input and speech output, reducing sensitivity to variations in instruction style. 
Since our augmentation strategy preserves the core semantics of the emotion labels with only enriched linguistic diversity, the model maintains stable emotional expressiveness.
All our experiments in this paper are conducted using the description-augmented EmoVoice-DB as the training dataset.


\vspace{-2mm}
\section{Discussion on Emotion Evaluation Metrics}

\begin{table}[ht]
 \vspace{-2mm}
\centering
\caption{Correlation between different metrics and human ratings measured by system-level and sentence-level Spearman's rank correlation coefficient($\rho$), and stability of LLM scoring measured by Standard Deviation(SD).}
\label{tab:metric}
 \vspace{-2mm}
\resizebox{\columnwidth}{!}{
\begin{tabular}{l l c c c}
\toprule
\multirow{2}{*}{Model} & \multirow{2}{*}{Metric} & \multicolumn{2}{c}{Spearman's $\rho$}  & \multirow{2}{*}{SD} \\
\cmidrule(lr){3-4}
      &        & System-level & Sentence-level &  \\
\midrule
emotion2vec & Emo\_Sim & 0.9429 & 0.4047 & / \\
GPT-4o-audio & Rating & 0.7714 & 0.2569 & 0.3286 \\
Gemini-2.0-flash & Rating & 0.9429 & 0.1960 & 0.2395 \\
\bottomrule
\end{tabular}
}
\vspace{-3mm}
\end{table}


Prior emotional TTS models are often trained with categorical emotion labels and evaluated with classification accuracy metrics. We adopt the finer metric emotion similarity to evalute emotion expressiveness and the consistency with detailed emotional instructions. 
However, while emotion similarity generally aligns with MOS (as shown in Table \ref{tab:mos} and Table \ref{tab:output_structure}) and reflects emotional expressiveness to certain extent, it is not sensitive enough to discern finer performance differences, particularly among closely matched models—such as different variants of EmoVoice model—where the numbers are nearly indistinguishable, indicating its inadequacy as a precise metric.

Therefore, we investigate the true evaluative capacity of emotion similarity. Besides, considering the versatile capabilities of multimodal LLMs, we also explore whether GPT-4o and Gemini can effectively assess emotional attributes of speech. 
We randomly select 10 description-text pairs from the test set from each emotion category excluding neutral category. Utilizing six distinct models—including GPT-4o series models, CosyVoice series models, and PromptTTS—we collect 360 audio samples of different qualities, which are then evaluated by 25 human raters. We filter a balanced set of audios with MOS scores evenly distributed across five levels (1 to 5), yielding a final set of 100 speech for further analysis.
Emotion similarites between these audios and their ground-truth counterparts from the test set are computed using emotion2vec. GPT-4o-audio and Gemini-2.0-Flash are instructed to assess their emotional expressiveness in a human-like manner, prompted with the test audio and a detailed subjective evaluation criteria. 
We rank similarity scores and model-generated ratings and compare them with human MOS rankings by calculating the Spearman's rank correlation coefficients to quantify the alignment between these metrics and human subjective judgments, denoted as sentence-level Spearman's $\rho$. Besides, we calculate the system-level Spearman's $\rho$ for emotion similarity based on Table \ref{tab:mos}, and prompt LLMs to rate the systems in it following similar procedure as the MOS evaluation.
Since LLMs exhibit variability when rating the same audio sample multiple times, we also calculate the standard deviation of their ratings to assess the stability and credibility of their judgments.

As shown in Table \ref{tab:metric}, system-level Spearman's $\rho$ are generally high showing minor ranking errors. However, all metrics have low sentence-level Spearman's $\rho$ below 50\%, indicating a generally weak alignment with human perception. In conclusion, emotion similarity offers certain reference value in assessing the overall quality of models, but lacks accuracy and sensitivity for fine-grained comparisons between individual audio samples. LLMs still lack the ability to effectively evaluate emotional speech. Besides, Gemini scores are relatively stable, whereas GPT-4o ratings fluctuate more.
The results highlight the urgent need for reliable, accurate, and fine-grained evaluation metrics that align with human perception, critical for advancing research and applications in emotion-related speech synthesis and analysis tasks.

\vspace{-2mm} 
\section{Conclusions}
In this work, we explore fine-grained emotion-controllable TTS task from the perspectives of \textit{model design}, \textit{data}, and \textit{evaluation metrics}. 
We propose EmoVoice, a novel emotion-controllable TTS model with freestyle text prompting based on LLMs, and EmoVoice-PP, a phoneme boost variant model that outputs phoneme tokens and audio tokens in parallel to enhance content consistency. We also compare different output structures, including parallel, sequential, and interleaved outputs.
Besides, we propose EmoVoice-DB, a high-quality 40-hour English emotion dataset with expressive speech and natural language emotion description labels.
EmoVoice achieves SOTA performance on the English EmoVoice-DB test set using only synthetic training data and on the Chinese Secap test set using in-house data.
We analyze the reliability of existing speech emotion evaluation metrics and explore using SOTA multimodal LLMs for evaluation. Both lack the precision and sensitivity required for accurate emotion evaluation, thus underscoring the need for fine-grained evaluation metrics that align with human perception.

\clearpage
\begin{acks}
This work was supported by the National Natural Science Foundation of China (No. U23B2018 and No. 62206171), Shanghai Municipal Science and Technology Major Project under Grant 2021SHZDZX0102 and Yangtze River Delta Science and Technology Innovation Community Joint Research Project (2024CSJGG01100).
\end{acks}

\bibliographystyle{ACM-Reference-Format}
\bibliography{sample-base}

\end{document}